\documentclass[11pt]{article}

\usepackage{cite}
\usepackage{helvet}
\usepackage{amsmath}
\usepackage{amssymb}
\usepackage{setspace}
\usepackage{graphicx}
\usepackage{empheq}
\usepackage{soul}
\usepackage{subcaption}
\usepackage{tabularx}
\usepackage{array}
\usepackage{float}
\usepackage{braket}
\usepackage[utf8]{inputenc}
\usepackage{url}
\usepackage{hyperref}
\usepackage{slashed}
\usepackage[symbol]{footmisc}
\usepackage{lineno} 

\def\be{\begin{equation}}
\def\ee{\end{equation}}

\usepackage[a4paper,top=3cm,bottom=3cm,left=2.5cm,right=2.5cm,bindingoffset=0mm]{geometry}

\begin{document}

 \begin{flushright}                                                    
 IPPP/20/33                                                                                                     
  \end{flushright}  


\begin{center}

{\Large \bf A new approach to modelling elastic and inelastic \\ \vspace{4mm}  photon-initiated production at the LHC: SuperChic 4}

\vspace*{1cm}
      
{\sc 
L.A.~Harland-Lang$^{1}$%
\footnote[1]{email: harland-lang@physics.ox.ac.uk}%
, M.~Tasevsky$^{2}$%
\footnote[2]{email: marek.tasevsky@cern.ch}%
, V.A.~Khoze$^{3,4}$%
\footnote[3]{email: v.a.khoze@durham.ac.uk}%
and M.G.~Ryskin$^{4}$%
\footnote[4]{email: ryskin@thd.pnpi.spb.ru}\\%
}
\vspace*{0.5cm}

{\small\sl
  $^1$Rudolf Peierls Centre, Beecroft Building, Parks Road, Oxford, OX1 3PU, UK

\vspace*{0.25cm} 

$^2$Institute of Physics, Czech Academy of Sciences, 
CS-18221 Prague 8, Czech Republic

\vspace*{0.25cm} 

$^3$IPPP, Department of Physics, University of Durham, 
Durham, DH1 3LE, UK

\vspace*{0.25cm}

$^4$Petersburg Nuclear Physics Institute, NRC ``Kurchatov Institute'', Gatchina,
St.~Petersburg, 188300, Russia
}

\vspace*{1cm}   

\begin{abstract}
\noindent  
We present the results of the new \texttt{SuperChic 4} Monte Carlo implementation of photon--initiated production in proton--proton collisions, considering as a first example the case of lepton pair production. This is based on the structure function calculation of the underlying process, and focusses on a complete account of the various contributing channels, including the case where a rapidity gap veto is imposed. We provide a careful treatment of the contributions where either (single dissociation), both (double dissociation) or neither (elastic) proton interacts inelastically and dissociates, and interface our results to {\tt Pythia} for showering and hadronization. The particle decay distribution from dissociation system, as well the survival probability for no additional proton--proton interactions, are both fully accounted for; these are essential for comparing to data where a rapidity gap veto is applied. We present detailed results for the impact of the veto requirement on the differential cross section, compare to and find good agreement with ATLAS 7 TeV data on semi--exclusive production, and provide a new precise evaluation of the background from semi--exclusive lepton pair production to SUSY particle production in compressed mass scenarios, which is found to be low.
\end{abstract}

\end{center}

\section{Introduction}

Photon--initiated (PI) particle production is a key ingredient in the LHC physics programme, playing a role in precision predictions for inclusive electroweak particle production~\cite{Manohar:2016nzj,Manohar:2017eqh,Harland-Lang:2019eai}, probes of Beyond the Standard Model (BSM) physics~\cite{Ohnemus:1993qw,Piotrzkowski:2000rx,Khoze:2001xm,N.Cartiglia:2015gve,Accomando:2016tah,Accomando:2016ehi,Khoze:2017igg,Baldenegro:2018hng,HarlandLang:2011ih,Beresford:2018pbt,Harland-Lang:2018hmi,Godunov:2019tmi,Godunov:2019jib,Dyndal:2020yen}, SM physics in the diffractive sector~\cite{Khoze:2001xm,N.Cartiglia:2015gve,Harland-Lang:2016apc,Luszczak:2018dfi,Forthomme:2018sxa,Goncalves:2020saa}, and in ultraperipheral heavy ion collisions~\cite{dEnterria:2013zqi,Beresford:2019gww,Harland-Lang:2018iur,Bruce:2018yzs,Coelho:2020saz}.

A unique feature of the PI channel in proton--proton collisions is that the colour singlet photon exchange naturally leads to exclusive events, where the photons are emitted elastically from the protons, which then remain intact. This is particularly relevant in the context of the dedicated forward proton detectors (FPDs) at the LHC, namely the AFP~\cite{AFP,Tasevsky:2015xya} and CT--PPS~\cite{CT-PPS}  FPDs which have been installed in association with both ATLAS and CMS, respectively. More generally, even if the initial--state photon is emitted inelastically, there is no colour flow as a result, and there is still a possibility for semi--exclusive events with rapidity gaps  in the final--state between the proton dissociation system(s) and the centrally produced object. Indeed,  a range of data on semi--exclusive lepton and $W$ boson pair production\footnote{In what follows we will for brevity use the term `semi--exclusive' to include both cases where the protons remain intact and where they undergo dissociation.} have been taken at the LHC by both ATLAS~\cite{Aad:2015bwa,Aaboud:2016dkv,Aaboud:2017oiq} and CMS~\cite{Chatrchyan:2011ci,Chatrchyan:2013akv,Khachatryan:2016mud}. A measurement of semi--exclusive lepton pair production has in addition been performed by CMS--TOTEM, with one proton tagged in the TOTEM detector~\cite{Cms:2018het}. As well as providing a test of the SM in this EW and diffractive sector, such studies can be used to constrain BSM physics, such as anomalous gauge couplings, as was done in~\cite{Chatrchyan:2013akv,Khachatryan:2016mud}.

In all of the above cases, events are selected by imposing a veto on additional tracks associated with the dilepton vertex (with further cuts imposed to reduce non--exclusive backgrounds), which effectively corresponds to the requirement of a rapidity gap in the central detector, for which no additional particle production is present. Such measurements are  necessarily semi--exclusive in nature: that is, the event sample will contain events with inelastic photon emission from the proton, but where the dissociation products lie outside the track veto region, as well as elastic events with intact protons in the final state. On the other hand, this rapidity gap topology is rather far from the standard inclusive case, where no rapidity requirement is imposed. The reasons for this are twofold: first, events where decay products from the proton dissociation system enter the veto region must be excluded, and second, there may be additional inelastic proton--proton QCD interactions (in other words, underlying event activity) that fill the gap region. The latter effect must be accounted for via the so--called `survival factor' probability of no additional proton--proton interactions~\cite{Harland-Lang:2014lxa,Harland-Lang:2015eqa}, 
while the former requires a fully differential treatment of the PI process, including a MC implementation such that the showering and hadronisation of the dissociation system may be accounted for.

In this paper, we present such a MC implementation, \texttt{SuperChic 4}, for the case of lepton pair production. In the inclusive channel, there has in recent years been significant progress in achieving in principle high precision prediction for photon--initiated production within collinear factorization~\cite{Manohar:2016nzj,Manohar:2017eqh} and directly in the structure function approach~\cite{Harland-Lang:2019eai}. We will make use of the approach of~\cite{Harland-Lang:2019eai} to provide a high precision prediction for the underlying PI process that is fully differential in the kinematics of the final--state protons and/or dissociation systems. This can then be interfaced to a general purpose MC for further showering/hadronization; in the current study we will make use of \texttt{Pythia 8.2}~\cite{Sjostrand:2014zea}. We in addition account for the survival factor, in a manner that take full account of the dependence of this quantity on the event kinematics and the specific channel (elastic or inelastic). \texttt{SuperChic 4} is the first generator of its kind to take account of all of these features, which are essential when providing results for semi--exclusive PI production, in a way that the individual elastic, SD and DD components can be included individually or in combination. As such we believe it will have multiple applications for LHC physics.

For example, in the experimental analyses discussed above, in order to isolate the purely exclusive process some subtraction of the contributions from lepton pair production in which the proton(s) undergo single dissociation (SD) and double dissociation (DD) has been imposed. The modelling of these processes has been based in part at least on the \texttt{LPAIR 4.0} MC~\cite{Vermaseren:1982cz,Baranov:1991yq}. However, this provides a rather outdated prediction for the process, and most significantly does not account for the soft survival factor. A further approach that is taken in the ATLAS analyses~\cite{Aad:2015bwa,Aaboud:2016dkv,Aaboud:2017oiq} in modelling the DD contribution involves the use of LO collinear factorization, in terms of a photon PDF, in combination with the multi--parton interaction (MPI) model of a general purpose MC to account for the survival factor. As well as effectively including the SD and elastic components in the cross section normalization, the MPI model will not account for the delicate dependence of the survival factor on the kinematics and spin structure of the PI process. \texttt{SuperChic 4} bypasses all of these issues, and provides a complete treatment of all channels, elastic and dissociative, and will be directly applicable in future analyses of this type. To demonstrate this, we will compare our results to the ATLAS measurement~\cite{Aad:2015bwa} at $\sqrt{s}=7$ TeV of semi--exclusive electron and muon pair production, finding an encouraging level of agreement.

As a further example, in~\cite{Harland-Lang:2018hmi}, the possibility of observing SUSY particles with EW couplings in a compressed mass scenario is discussed, with the signal corresponding to relatively low $p_\perp$ leptons from the decay of the SUSY particles and FPD tags for the elastic protons. A potentially important background here corresponds to the SD and DD production of lepton pairs at relatively low mass, with a proton produced from the decay of the dissociation system registered in the FPD. As we will demonstrate, our new implementation provides all the necessary tools to evaluate this background.

In general, though we focus on the case of lepton pair production here, we emphasise that this MC can be readily extended to the PI production of other SM and BSM states. Indeed, in many cases a useful method to effectively isolate the PI contribution, with its unique colour singlet initiated topology, is to impose rapidity veto/isolation requirements on the final state. However, in order to account for this theoretically one must always go beyond the standard inclusive framework, and account for elastic and inelastic photon emission differentially, and in a way that accounts fully for the gap survival probability. The goal of the current paper is to provide such a MC implementation, in order to compare directly with data selected in this way.

Finally, though this is not the principle focus of the current paper, we note that the MC can be used to provide a high precision prediction for the fully inclusive PI cross section for lepton pair production, as given by the structure function calculation.

The outline of our paper is as follows. In Section~\ref{sec:SF} we summarise the key elements of the structure function calculation of the underlying PI process. In Section~\ref{sec:MC} we describe how this calculation is implemented in the MC and in particular can be interfaced to \texttt{Pythia}. In Section~\ref{sec:S2} we describe how the survival factor is included. In Section~\ref{sec:rapgap} we present some general results for the impact of imposing a rapidity gap veto on the relative contributions from the elastic, SD and DD channels  to PI production. In Section~\ref{sec:ATLAS} we compare predictions for the lepton acoplanarity distribution to the ATLAS data~\cite{Aad:2015bwa}. In Section~\ref{sec:DM} we evaluate the background from semi--exclusive lepton pair production to slepton pair production. Finally, in Section~\ref{sec:conc} we conclude and consider future extensions of our work.

\section{Theoretical Ingredients}

\subsection{Structure Function Calculation}\label{sec:SF}

The calculation of the inclusive photon--initiated cross section proceeds as described in~\cite{Harland-Lang:2019eai}. We will briefly summarise the key elements here, but refer the reader to this work for further detailed discussion and references. For the production cross section we can write
 \be\label{eq:sighhf}
 \sigma_{pp} = \frac{1}{2s}  \int  {\rm d}x_1 {\rm d}x_2\,{\rm d}^2 q_{1_\perp}{\rm d}^2 q_{2_\perp
} {\rm d \Gamma} \,\alpha(Q_1^2)\alpha(Q_2^2) \frac{1}{\tilde{\beta}}\frac{\rho_1^{\mu\mu'}\rho_2^{\nu\nu'} M^*_{\mu'\nu'}M_{\mu\nu}}{q_1^2q_2^2}\delta^{(4)}(q_1+q_2 - k)\;,
 \ee
 where $x_i$ and $q_{i\perp}$ are the photon momentum fractions (see~\cite{Harland-Lang:2019zur} for precise definitions) and transverse momenta, respectively.  Here the photons have momenta $q_{1,2}$, with $q_{1,2}^2 = -Q_{1,2}^2$, and we consider the production of a system of 4--momentum $k = q_1 + q_2 = \sum_{j=1}^N k_j$ of $N$ particles, where ${\rm d}\Gamma = \prod_{j=1}^N {\rm d}^3 k_j / 2 E_j (2\pi)^3$ is the standard phase space volume. $M^{\mu\nu}$ corresponds to the $\gamma\gamma \to X(k)$ production amplitude, with arbitrary photon virtualities. $\tilde{\beta}$ is as defined in~\cite{Harland-Lang:2019zur}.

In the above expression, $\rho$ is the density matrix of the virtual photon, which is given in terms of the well known proton structure functions:
 \be\label{eq:rho}
 \rho_i^{\alpha\beta}=2\int \frac{{\rm d}M_i^2}{Q_i^2}  \bigg[-\left(g^{\alpha\beta}+\frac{q_i^\alpha q_i^\beta}{Q_i^2}\right) F_1(x_{B,i},Q_i^2)+ \frac{(2p_i^\alpha-\frac{q_i^\alpha}{x_{B,i}})(2p_i^\beta-\frac{q_i^\beta}{x_{B,i}})}{Q_i^2}\frac{ x_{B,i} }{2}F_2(x_{B,i},Q_i^2)\bigg]\;,
 \ee
where $x_{B,i} = Q^2_i/(Q_i^2 + M_{i}^2 - m_p^2)$ for a hadronic system of mass $M_i$, and the integral over $M_i^2$ is understood as being performed simultaneously with the other phase space integrals. This corresponds to the general Lorentz--covariant expression that can be written down for the photon--hadron vertex, and Eq.~\eqref{eq:sighhf}, combined with a suitable input for the proton structure functions, represents the complete result we need in order to calculate the corresponding photon--initiated cross section in proton--proton collisions. 

The input for the proton structure functions comes from noting that the same density matrix $\rho$ appears in the cross section for lepton--proton scattering. One can therefore make use of the wealth of data for this process to constrain the structure functions, and hence the photon--initiated cross section, to high precision. In more detail, the structure function receives contributions from: elastic photon emission, for which we use the A1 collaboration~\cite{Bernauer:2013tpr}  fit to the elastic proton form factors;  CLAS data on inelastic  structure functions in the resonance $W^2 < 3.5$ ${\rm GeV}^2$ region, primarily concentrated at lower $Q^2$ due to the $W^2$ kinematic requirement; the HERMES fit~\cite{Airapetian:2011nu} to the inelastic low $Q^2 < 1$ ${\rm GeV}^2$ structure functions in the continuum $W^2 > 3.5$ ${\rm GeV}^2$ region; inelastic high $Q^2 > 1$ ${\rm GeV}^2$ structure functions for which the pQCD prediction in combination with PDFs determined from a global fit  provide the strongest constraint (we take the ZM--VFNS at NNLO in QCD  predictions for the structure functions as implemented in~\texttt{APFEL}~\cite{Bertone:2013vaa}, with the \texttt{MMHT2015qed\_nnlo} PDFs throughout, though in the MC the PDF can be set by the user). The inputs we take are as discussed in the MMHT15 photon PDF determination~\cite{Harland-Lang:2019pla}, which itself is closely based on that described in~\cite{Manohar:2016nzj,Manohar:2017eqh} for the \texttt{LUXqed} set. However, we note that our calculation makes no explicit reference to the partonic content of the proton itself, and as discussed in detail in~\cite{Harland-Lang:2019eai} provides by construction a more precise prediction than the result within collinear factorization that uses such a photon PDF.

\subsection{Monte Carlo implementation and treatment of proton dissociation}\label{sec:MC}

The expression \eqref{eq:sighhf}, in combination with the structure function inputs described above, is immediately amenable to a MC implementation of both elastic and inelastic photon--initiated production, by simply applying the  elastic or inelastic structure function at the corresponding vertex.  In particular, we can generate a fully differential final--state in terms of not just the centrally produced system, but the squared photon virtualities $Q^2_i$ and the invariant masses $M_{i}$ of the proton dissociation systems, for the case of inelastic emission, while for elastic emission the corresponding structure functions are simply $\propto \delta(x_{B,i}-1)$, implying $M_i=m_p$ as expected.

On the other hand, while the structure function calculation provides a precise prediction for the 4-momentum of the outgoing proton system as well as the initiating photon, to make contact with data we must also account for the decay of this system, which will as usual involve parton showering and subsequent hadronization. This is a non--trivial problem: in principle, in for example the resonance region we should take care to account for the appropriate branchings of the proton excitations, while in the high $Q^2$ region the parton shower should match that coming from the parton--level NNLO prediction (itself an open problem) for the proton structure function. 

We will take a generalised approach, which aims to capture the key physical expectations for proton dissociation via photon emission. Namely, the amount of particle production should be driven by the scale of the photon $Q^2$ and the invariant mass $M_i$ of the dissociation system, and should occur essentially independently of any dissociation on the other proton side, being colour disconnected from it. To achieve this, we have interfaced the appropriately formatted unweighted Les Houches events (LHE) to \texttt{PYTHIA 8.2}, with a suitable choice of run parameters. As this general purpose MC is set up to read in parton--level events, with collinear initiating partons, we simply map the kinematics of the $p \to \gamma + X$  process onto the parton--level $q\to q + \gamma$ process, where the photon 4--momenta are left unchanged, and the momentum fraction of the collinear initial--state quark is set so as to reproduce the appropriate $x_{B,i}$, and hence invariant mass $M_i$ of the proton final--state, provided by the structure function calculation. This may then be passed to \texttt{PYTHIA} for showering/hadronization. For concreteness we assign the collinear initiator to be an up quark, but the final result should not be sensitive to this choice. Indeed, a similar approach has been taken in~\cite{Forthomme:2018sxa}, where this point was verified explicitly.

For the  \texttt{PYTHIA} settings, we first of all set \texttt{PartonLevel:MPI=off}, as we only consider those events with no addition MPI, as accounted for via our calculation of the survival factor. We also use the dipole recoil scheme discussed in~\cite{Cabouat:2017rzi}, which is specifically designed for cases where there is no colour flow between the two initiating protons (or in the parton--level LHE, quarks), as is the case here. Taking the default global recoil scheme leads to a significant overproduction of particles in the central region. A similar effect has recently been observed in~\cite{Jager:2020hkz} for the case of Higgs boson production via vector boson fusion, for which a significant enhancement of jet production at central rapidities was seen, and it was verified by comparison with the NLO calculation of Higgs boson production + 3 jets at NLO+PS accuracy that this effect was unphysical. As recommended in the \texttt{PYTHIA} user manual, we take \texttt{SpaceShower:pTmaxMatch = 2 }, in order to fill the whole phase space with the parton shower, but we set \texttt{SpaceShower:pTdampMatch=1} to damp emission when it is above the scale \texttt{SCALUP} in the LHE, which we set to the maximum of the two photon $\sqrt{ Q_{i}^2}$; in practice, this latter option is found to have little effect on the results. We in addition set \texttt{BeamRemnants:primordialKT = off}, as we wish to keep the initiating quark completely collinear to fully match the kinematics from the structure function calculation. 

Finally, for an elastic proton vertex, we must include the initiating photon in the event in order for \texttt{Pythia} to process it correctly. For the case of SD, this requires the kinematics to be modified in order to keep the elastic photon collinear and on--shell. This is achieved by setting the photon transverse momentum to zero in the event (but not in the cross section calculation), and keeping the momentum fraction fixed.  We note this is only a technical necessity in order for \texttt{Pythia} to correctly handle the event, for the specific case of SD production, which features one elastic and one inelastic vertex. In particular we set \texttt{SpaceShower:QEDshowerByQ = off}, such that there is no back evolution from the photon, consistent with this being an elastic emission. Treating the initiating photon as on--shell in the event kinematics is of course an approximation to the true result, but for most purposes is a very good one. 

\subsection{Rapidity gaps: inclusion of survival factor}\label{sec:S2}

The results in the previous sections allow for a particle--level prediction of inclusive photon--initiated production, calculated with the structure function approach. However, as discussed in the Introduction, experimentally we may be interested in semi--exclusive production, that is the central production of EW objects (leptons, $W$ pairs, sleptons, electroweakinos, ALPs...) with no additional particles produced in the final state. In this case we need to include in addition the probability of no additional particle production due to soft proton--proton interactions (i.e. underlying event activity), known as the survival factor, see~\cite{Harland-Lang:2014lxa,Harland-Lang:2015eqa} for reviews.

We note that in all our results, and in the MC implementation, we always include this probability of no additional inelastic proton--proton interactions, however strictly speaking the result of such an interaction might still pass the experimental rapidity veto, i.e. the additional underlying event activity could lie entirely outside the rapidity veto region. However, we recall that any additional inelastic proton interactions will generate colour flow between the colliding protons, and as is well known the probability of producing such a gap from this non--diffractive interaction is exponentially suppressed by the size of the rapidity veto. Therefore for a reasonable veto region the contribution from inelastic events which pass the experimental veto should be small, with the precise amount depending on both the gap size and the $p_\perp$ threshold, see e.g.~\cite{Khoze:2010by} for a specific calculation. 
One could  estimate the size of such an effect by a suitable analysis of the particle distribution generated by the model of MPI in a general purpose MC. More precisely, the value of the uncorrected survival factor should be generated using our MC, which as we discuss provides a full account of the kinematic dependence of this quantity and its variation between the different dissociative channels. From this, we could derive a contribution from the fraction $\sim 1 -  S^2$ of non--diffractive events for which additional proton--proton interactions occur, but where a rapidity gap is still present within the experimental acceptance. Roughly speaking this will be at the $\sim \%$ level, and hence as for elastic and SD interactions (see below) we have $S^2 \sim 0.9$, the overall correction to $S^2$  will be extremely small, at the per mille level. For DD interactions on the other hand, where $S^2 \sim 0.1$, we may expect the corrected survival factor to be larger by a few percent. On the other hand, this is precisely the region where the theoretical uncertainty on the original survival factor itself is larger, certainly within this percent level correction.

It should be emphasised that  the impact of survival effects depends sensitively on the subprocess, through the specific proton impact parameter dependence. For example, it has been known for a long time\cite{Khoze:2001xm,Khoze:2002dc} that for purely elastic photon--initiated production, the low virtuality photon kinematics corresponds to a relatively large impact parameter separation between the colliding protons, and hence a smaller probability of additional proton--proton interactions, that is an average survival factor quite close to unity. Experimental evidence of this in proton--proton collisions has been seen in e.g.~\cite{Aaboud:2017oiq} while the same effect occurs in heavy ion collisions, for which the colliding ions are ultra--peripheral and the SM prediction for e.g. light--by--light scattering agrees well with the data~\cite{Aaboud:2017bwk,Sirunyan:2018fhl,Aad:2019ock} after including relatively mild survival effects.

To account for the above effects we follow the approach described in~\cite{Harland-Lang:2015cta,Harland-Lang:2016apc}. We recap the relevant issues here, but refer the reader to these for more details. One can write the averaged survival factor as
\begin{equation}\label{seikav}
\langle S_{\rm eik}^2\rangle= \frac{\int {\rm d}^2 q_{1_\perp}\,{\rm d}^2 q_{2_\perp}\,|T(q_{1_\perp},q_{2_\perp})+T^{\rm res}(q_{1_\perp},q_{2_\perp})|^2}{\int {\rm d}^2q_{1_\perp}\,{\rm d}^2q_{2_\perp}\,|T(q_{1_\perp},q_{2_\perp})|^2}\;,
\end{equation}
where $q_{i_\perp}$ are as before the photon transverse momenta. Here $T(q_{1_\perp},q_{2_\perp})$ is the amplitude  corresponding to the cross section given in \eqref{eq:sighhf}, while the so--called `screened' amplitude defines the effect of potential proton--proton interactions, and is given in terms of the original amplitude $T$ and the elastic proton--proton scattering amplitude, with\footnote{Strictly speaking, this expression corresponds to the so--called the single--channel approximation. We show this for the sake of clarity but note that in actual calculations we use the two--channel approach described in~\cite{Khoze:2013dha}.}
\begin{equation}\label{skt}
T^{\rm res}({q}_{1_\perp},{q}_{2_\perp}) = \frac{i}{s} \int\frac{{\rm d}^2  {k}_\perp}{8\pi^2} \;T_{\rm el}(k_\perp^2) \;T({q}_{1_\perp}',{q}_{2_\perp}')\;,
\end{equation}
where $q_{1_\perp}=q_{1_\perp}'+k_\perp$ and $q_{2_\perp}'=q_{1_\perp}-k_\perp$. The elastic amplitude itself can be written in impact parameter space in terms of the probability $\exp(-\Omega(s,b_\perp))$, that no inelastic proton--proton scattering occurs at impact parameter $b_\perp$ between the colliding protons, where $\Omega(s,b_\perp)$ is known as the proton opacity. As $b_\perp$ decreases, additional short--range QCD interactions become more likely, and this probability tends to zero. Through this, the dependence of the survival factor \eqref{seikav} on the particular form of the amplitude $T$ in photon $q_{i\perp}$ and hence proton impact parameter space, enters.

In any calculation of the survival factor we are therefore interested in extracting this vector $q_{i\perp}$ dependence of the production amplitude as accurately as possible. Now, a potential issue is that the structure function result \eqref{eq:sighhf} is given only at the cross section, and not amplitude level. However, as discussed in~\cite{Harland-Lang:2015cta}, at lower photon $q_\perp$ one can isolate the dominant contribution at the amplitude level from \eqref{eq:rho}, by noting that that the first term $\sim F_1$ is suppressed by $\sim x_i^2/x_{B,i}^2$ with respect to the second term $\sim F_2$. For this term we can identify the contribution at the amplitude level
\be\label{eq:Tprop}
T \propto p_{1}^\mu p_{2}^\nu M_{\mu \nu} \approx \frac{q_{1\perp}^\mu q_{2 \perp}^\nu}{x_1 x_2} M_{\mu \nu} \;,
\ee
where the second relation comes from the high energy expansion $q_i = x_i p_i + q_{i\perp}$ and the gauge invariance of $M$. In other words, we apply the usual eikonal approximation for the $p\to X + \gamma$ vertex. We therefore make use of this relation, with the subleading terms $\propto F_1$ included incoherently in the amplitude, as in~\cite{Harland-Lang:2015cta}. Although such a result applies equally well for the parton--level diagrams entering a pQCD calculation of the structure functions as for elastic photon emission, nonetheless in the  $Q^2>Q_0^2$ region of inelastic production, we are applying the result within collinear factorization in terms of proton quark/gluon PDFs that are defined at the cross section and not amplitude level. This may be a cause for concern, and hence at high enough $Q^2$ we can take an alternative approach. This comes from observing that in the $Q^2 \gtrsim Q_0^2$ regime we have $q_{i_\perp}' \approx q_{i_\perp}$ and the amplitude $T$ factorizes from the integral in \eqref{skt}. In such a case it only ever appears at the $|T|^2$ level in \eqref{seikav}. More precisely, one can model the $k_\perp$ dependence entering the integral in \eqref{skt} from the inelastic cross section with reference to the dependence in the case of the `generalized' PDFs~\cite{Diehl:2003ny,Belitsky:2005qn}, which allowed for such a non--zero momentum transfer, in terms of the proton Dirac form factor $F_1$.
This complete factorization only applies for the case that both emitted photons are emitted inelastically with $Q^2 \gtrsim Q_0^2$, but in the mixed case where one photon is emitted elastically or inelastically but at low $Q^2$, a similar procedure can be performed. We emphasise again that the above discussion only serves as a summary of the key element of the calculation, and refer the reader to~\cite{Harland-Lang:2016apc} for full details.

\begin{figure}[h]
\begin{center}
\includegraphics[scale=0.64]{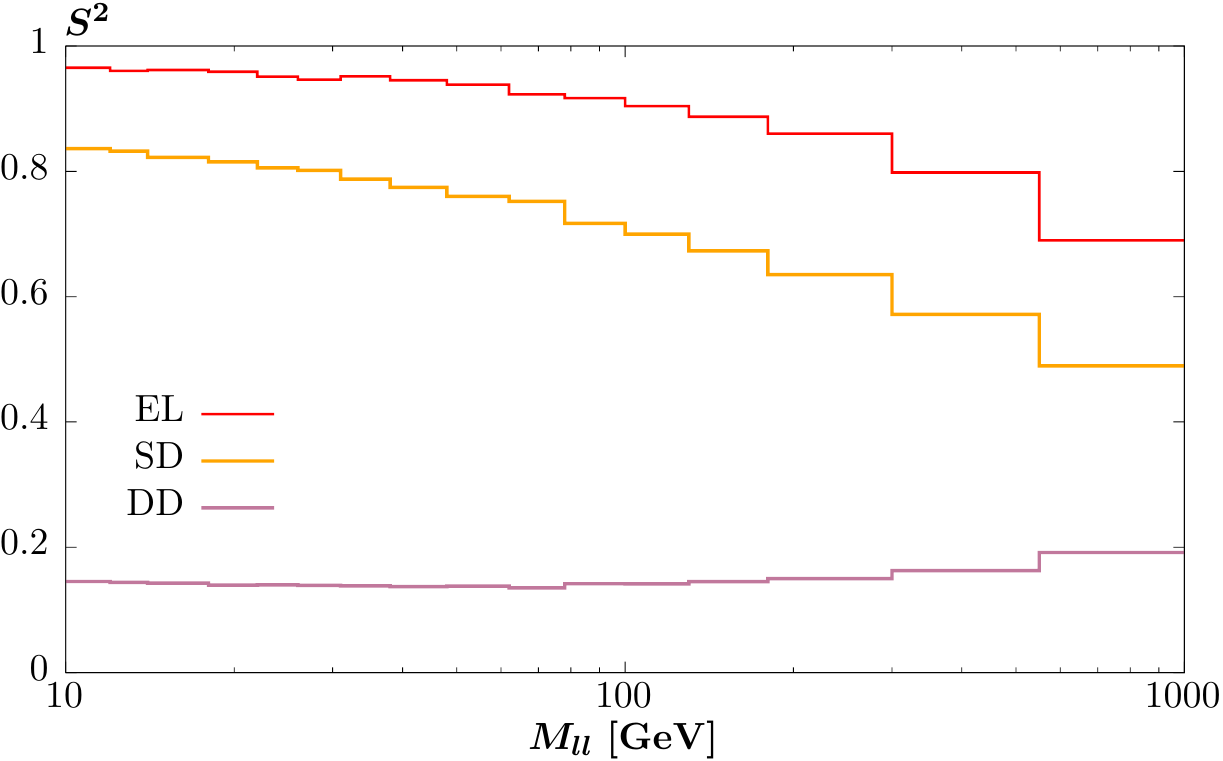}
\includegraphics[scale=0.64]{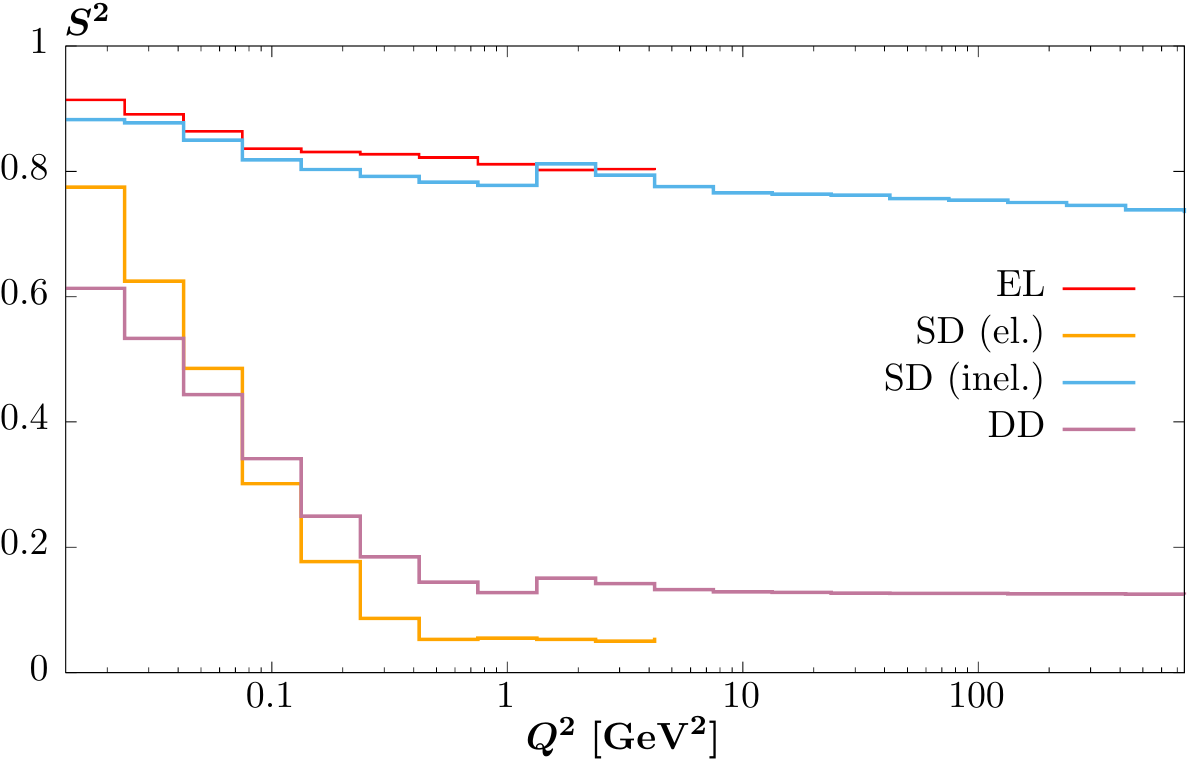}
\includegraphics[scale=0.64]{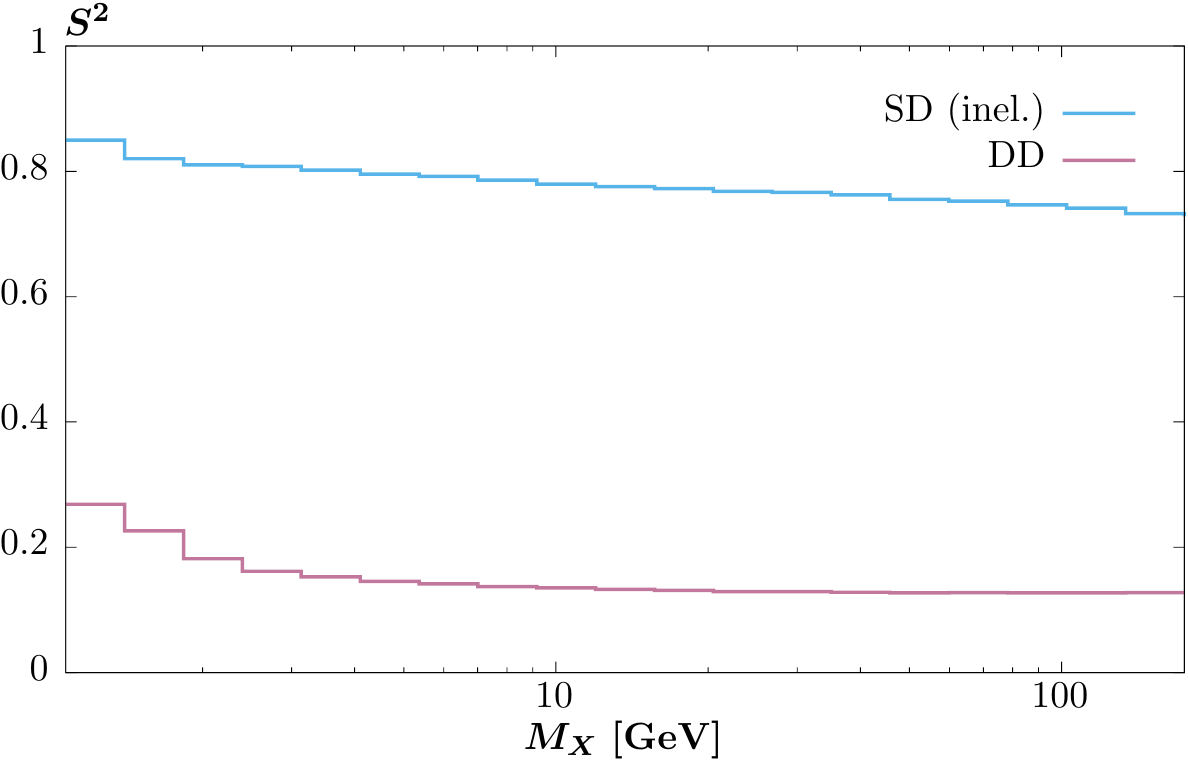}
\includegraphics[scale=0.64]{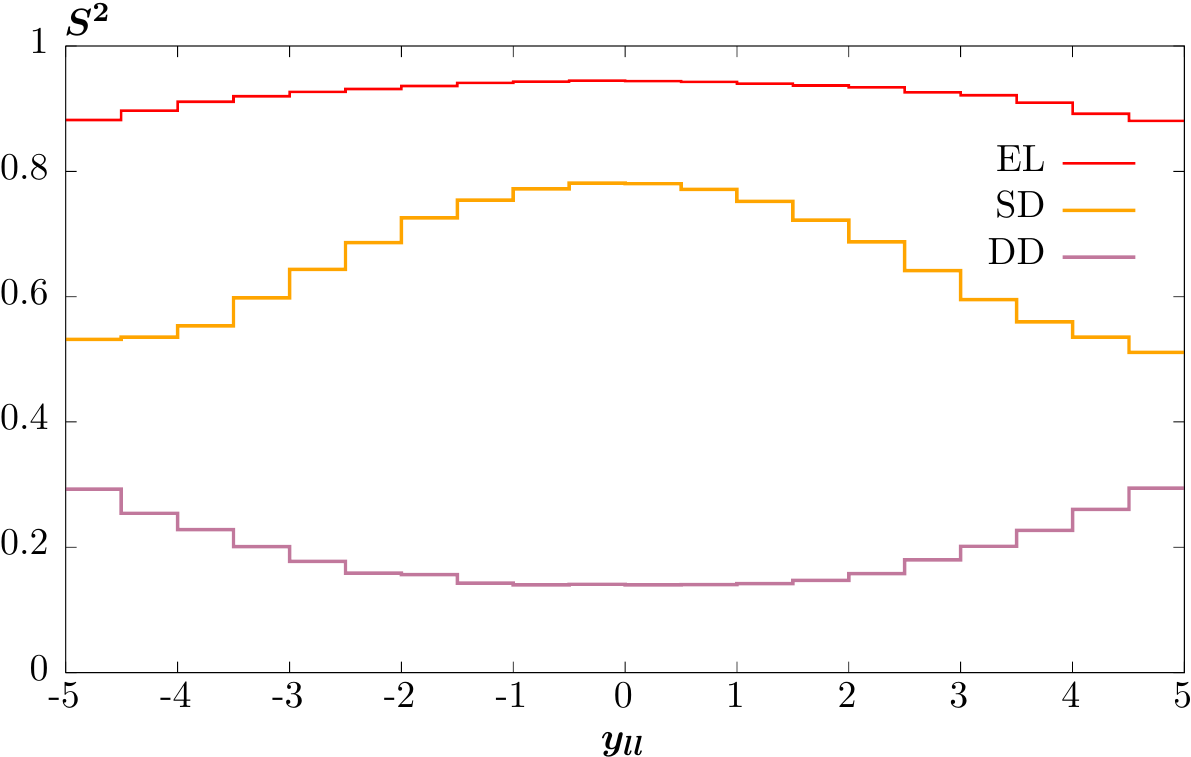}
\caption{Soft survival factor for lepton pair production as a function of the invariant mass, $M_{ll}$, of the dilepton system, the photon $Q^2$, the invariant mass of the dissociation system, $M_{\rm X}$, and the dilepton rapidity, $y_{ll}$. Results are given for elastic, SD (with the elastic or inelastic vertex indicated where relevant) and DD, and correspond to muon pair production with $\sqrt{s}=13$ TeV and lepton $p_\perp^l> 10$ GeV, $|\eta_l|<2.5$, though the results are largely insensitive to this precise choice of cuts, and lepton mass effects are negligible in this region. For the invariant mass distribution we impose a lower cut of $p_\perp^l > 1$ GeV  in order to reduce kinematic effects at low masses. For the elastic and SD (elastic) $Q^2$ distributions, the plots are cutoff when the effect of limited statistics due to the sharply falling form factors begins to dominate.}
\label{fig:S2}
\end{center}
\end{figure} 

We note that from the discussion above we expect a transition in the appropriate calculation of the survival factors to occur in the $Q^2 \sim Q_0^2$ region, however clearly there is some ambiguity as to the precise scale below which one can/should work at the amplitude level and above which at the cross section level. In the MC we therefore include both evaluations for each point in phase space, with a smooth interpolation performed between the two regimes around $Q^2 = Q_0^2=1$ ${\rm GeV}^2$, such that the relevant amplitude (cross section) level calculation dominates in the $Q^2 \ll Q_0^2$ ($Q^2 \gg Q_0^2$) regime.

What are the consequences of the above calculation for inelastic photon--initiated production? As discussed above, the survival factor depends on the photon virtuality through its effect on the impact parameter of the colliding protons. In particular, we will expect that for inelastic photon emission, the larger photon $Q^2$ will result in a smaller survival factor and hence a suppressed rate, with this effect becoming more significant as the photon $Q^2$ increases. Such an effect was indeed observed in the results of~\cite{Harland-Lang:2016apc}.

In Fig.~\ref{fig:S2} we show results for the soft survival factor as function of various kinematic variables for muon pair production at $\sqrt{s}=13$ TeV with $|p_\perp^l|> 10$ GeV (1 GeV for the dilepton invariant mass distribution), $|\eta_l|<2.5$, though the results are largely insensitive to this precise choice of cuts. 
Considering first the dependence on the dilepton invariant mass, we can see that broadly there is a large difference in the magnitude of the survival factor between the DD and elastic/SD cases, with the former being significantly smaller. This in line with the expectations above, being in particular driven by the fact that in the DD case the photon $Q^2$ is generally much higher, and so the collision is less peripheral; the most peripheral elastic interaction has the highest survival factor. We can also see that as the invariant mass increases, the survival factor decreases, due to effect of the kinematic requirement for producing an on-shell proton at the elastic vertex for larger photon momentum fractions, which implies a larger photon $Q^2$, see~\cite{Harland-Lang:2015cta}. For the DD case the survival instead increases somewhat, due to the smaller phase space in photon $Q^2$ at the highest $M_{ll}$ values.

The photon $Q^2$ distribution, which while not individually an observable (with the exception of the elastic case with proton tagging) is nonetheless an illustrative demonstration of the underlying physics and is plotted as well. We show the inclusive binned photon $Q^2$  from both vertices in the elastic and DD cases, while for SD we distinguish between the elastic and inelastic photon vertices. For the elastic and SD (inelastic) cases we observe a mild reduction with increasing $Q^2$, due to the fact that the average photon $Q^2$ from the other, elastic, vertex is always low, leading to a peripheral interaction and higher survival factor. In contrast, for the SD (elastic) case we observe a steep fall with $Q^2$, as now the other inelastic, vertex has a relatively large $Q^2$ and hence the larger $Q^2$ region on the figure corresponds to a less peripheral interaction. Note that at lower $Q^2$, below the limit of the plot, the survival factor in the SD (elastic) case becomes larger than in the SD (inelastic) case, as required by the fact that these should integrate to the same average value. For the DD case a similar fall off is evident, with the result being constant at large enough $Q^2$. We note that this effect occurs by construction in our approach: in particular, as discussed below \eqref{eq:Tprop} at larger $Q^2$ we assume that the substructure of the squared $p\to \gamma^* X$ vertex can be factorized entirely from the survival factor calculation, with the $k_\perp$ dependence (corresponding to the impact parameter profile of the interaction) given in terms of the proton Dirac form factor. Thus we see a fall off as we interpolate between the lower $Q^2$ region, where we treat the kinematics differentially and work at the amplitude level when calculating the survival factor, and the constant tail, where we apply this factorized approach.
Strictly speaking, this factorized approach is derived by considering the cross section integrated over the photon $Q^2>Q_0^2$ and over the mass of the dissociation system. A more detailed, though necessarily rather model--dependent approach would account for this region differentially in these variables. However, we can see that in this region the survival factor is already rather low, and we consider such an effect to be within the uncertainties of the overall approach, though potentially worth further investigation in the future. We in addition note a slight kink at the transition region $Q^2 \approx Q_0^2$ in both the DD and SD (inel.) cases, which is a feature of the fact that the calculation of the survival factor in the higher $Q^2$ region corresponds to a slightly larger value of the $S^2$ in this region. A more complete, and smoother treatment of this region again may warrant further investigation, but is certainly within the uncertainty of the overall approach. 

We also show the dependence of the survival factor on the invariant mass of the dissociation system in the SD and DD cases and the dilepton rapidity in all cases. While the former variable is again not an observable on its own it highlights some of the underlying physics. We can see that for larger masses, where the interaction tends to be less peripheral, the survival factor becomes smaller. For the rapidity distribution, we can see in the elastic and SD cases a clear trend for the survival factor to decrease at larger rapidities. This effect is driven by the same kinematic requirement for the on--shell elastic proton as in the case of the invariant distribution above. For DD, where neither proton remains intact and hence this requirement is absent, the opposite trend is observed and the survival factor is found to increase at forward rapidities, which is driven by the smaller phase space for dissociation and hence lower average photon $Q^2$.

\section{Results}

\subsection{Impact of rapidity veto}\label{sec:rapgap}

\begin{figure}[h]
\begin{center}
\includegraphics[scale=0.26]{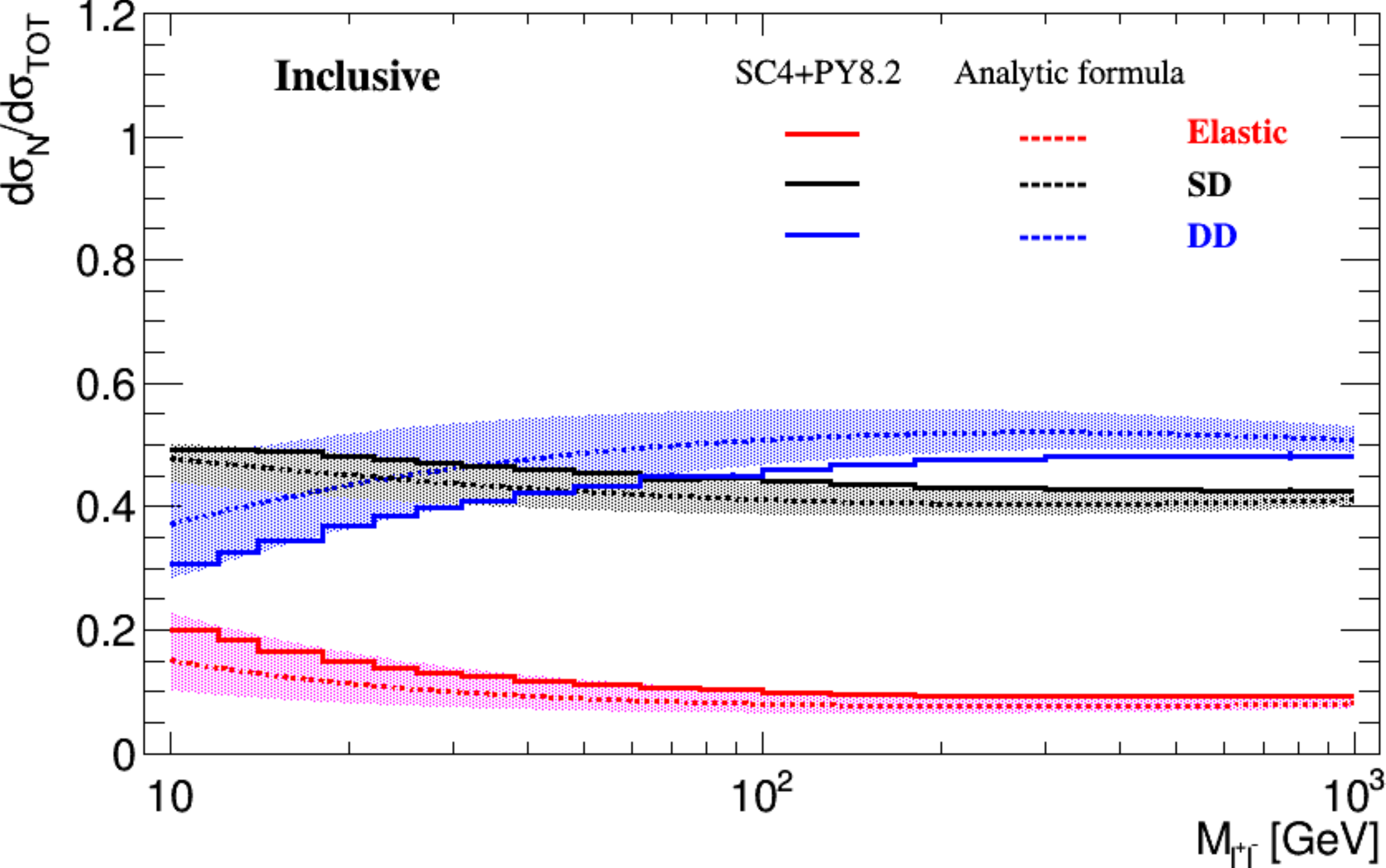}
\includegraphics[scale=0.26]{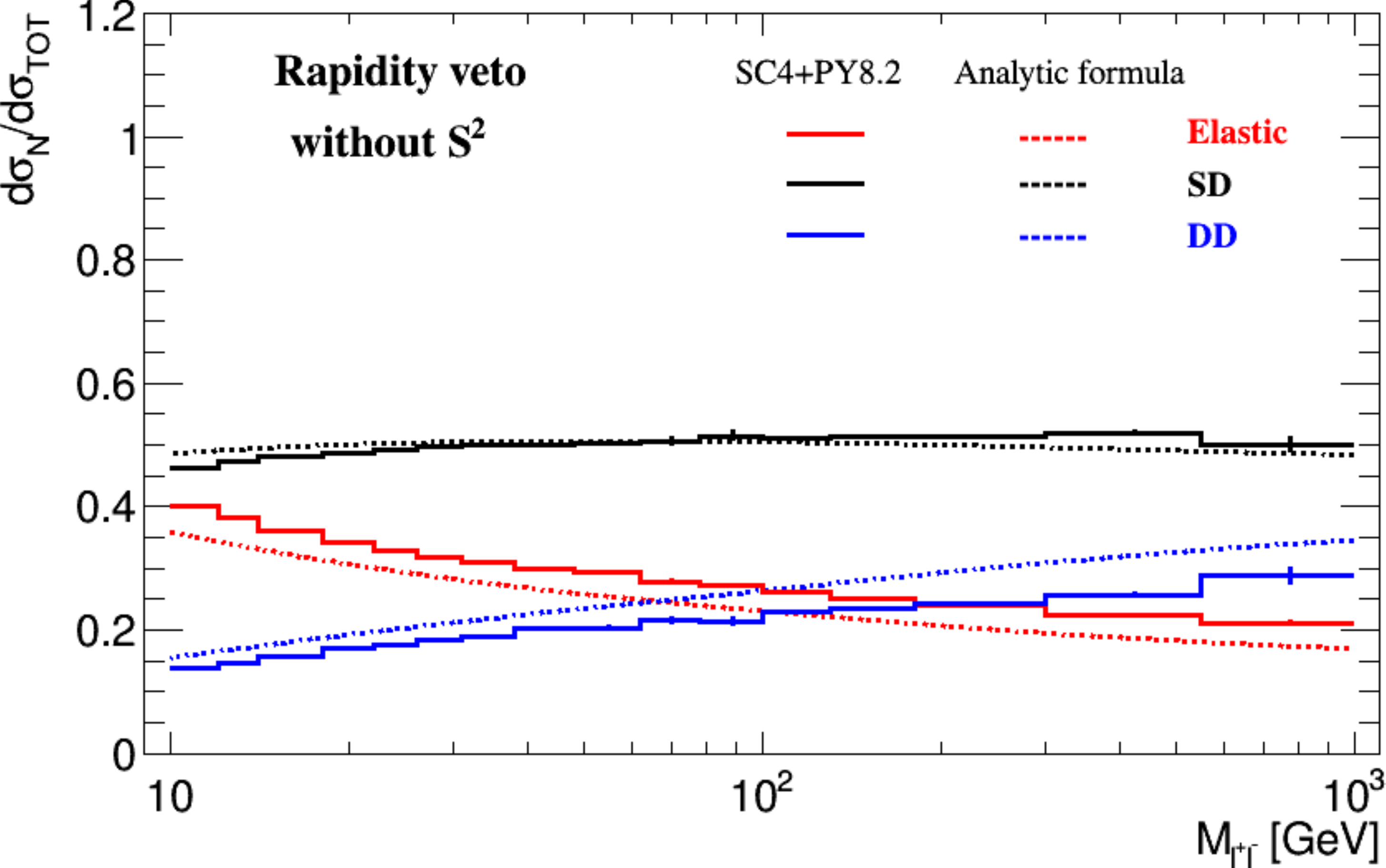}
\includegraphics[scale=0.26]{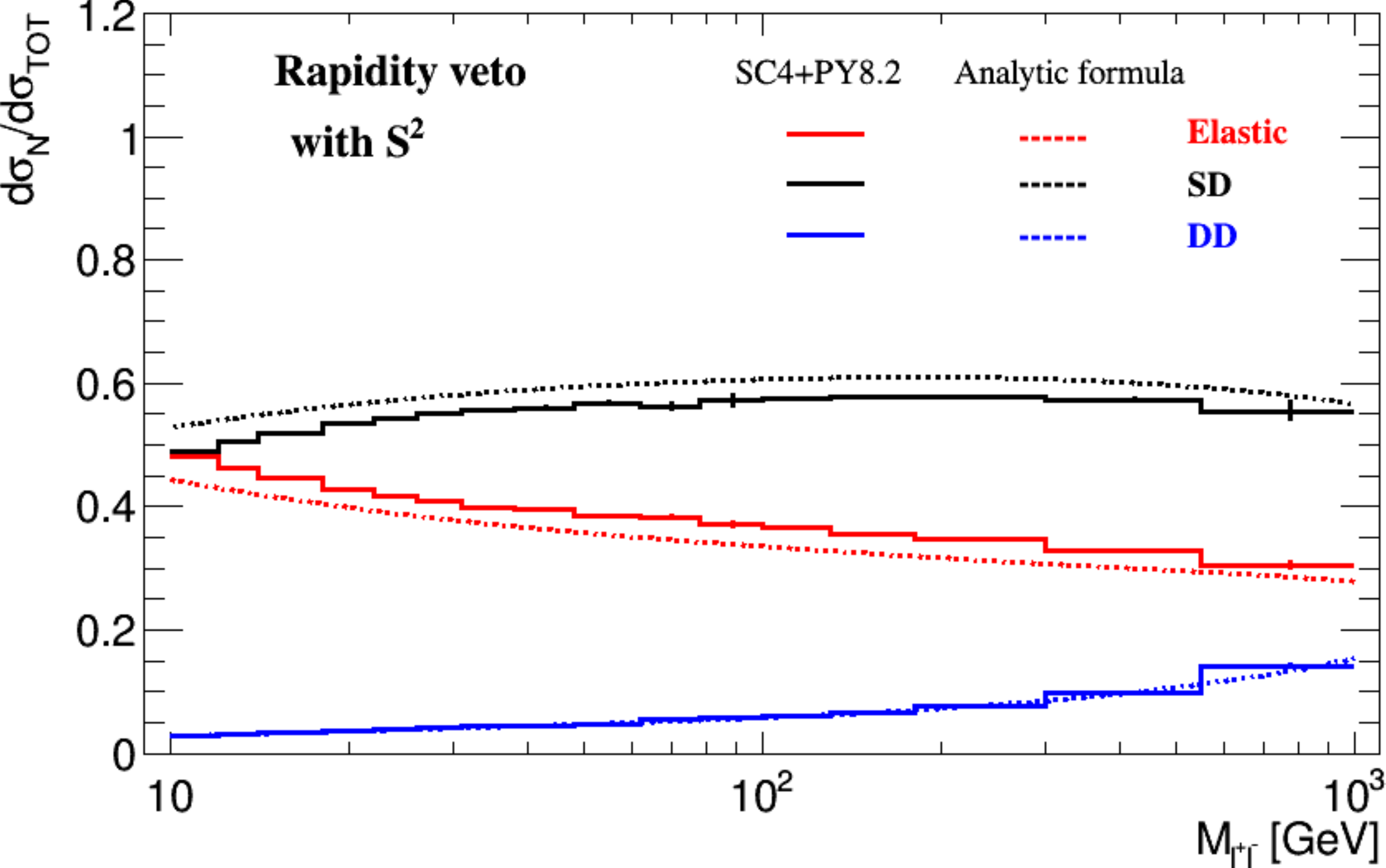}
\caption{Fractional cross section for elastic, SD and DD muon pair production, as a function of the dilepton invariant mass, at $\sqrt{s}=13$ TeV, for $p_\perp^l > 1$ GeV and $|\eta_l|<2.5$. The very low $p_\perp$ cut is chosen purely for display purposes, that is to reduce the impact of any kinematic effects of cuts at lower invariant mass. Results are shown for the inclusive case, after imposing a rapidity veto on additional particle production from the proton dissociation system alone in the $|\eta|<4.4$ region, and after including the survival factor from the MC. The analytic predictions of~\cite{Harland-Lang:2016apc}, based on modified DGLAP evolution, are shown as dashed lines, with factor of $ \mu_{F,R}= \mu_0\,{}^\times_\div 2$ (with $\mu_0=M_{ll}$) scale variation bands given in the inclusive case.}
\label{fig:veto}
\end{center}
\end{figure} 

We first consider the impact of the rapidity veto and survival factor on the relative contributions from elastic, SD and DD production, shown in Fig.~\ref{fig:veto}. For concreteness, we consider a veto in the $|\eta | < 4.4$ region on all particles, for ease of comparison with the analytic predictions of~\cite{Harland-Lang:2016apc}. We have checked that the result is very similar if we instead cut on charged particles, as would be done experimentally for a track veto. The solid histograms indicate the result of running \texttt{SuperChic 4}, interfaced to \texttt{Pythia 8.2} for showering and hadronisation of the proton dissociation system, as described in the previous sections. In the inclusive case, we can see that the fractional elastic component is the lowest, in the 10--20\% region, and decreases with mass, due to the larger phase space for proton dissociation that opens up in the SD and DD cases. Once the rapidity veto is imposed, we can see that the fraction of elastic events increases to the 20--40\% level; the veto will remove some SD and DD events where the dissociation products enter the veto region, but will leave the elastic events unaffected. For the same reason, we can see that the relative fraction of DD events is particularly affected, while the impact on the SD case, where only one proton interacts inelastically, is milder. The basic trend, with the fraction of elastic events decreasing with mass, is preserved. Once the survival factor is included, the fraction of DD events is further reduced, due to the effect observed in Fig.~\ref{fig:S2}, whereby the survival factor is significantly lower in this case. In summary, whereas for inclusive photon--initiated production the SD and DD are expected to be largely dominant, once one considers a more exclusive observable and imposes a rapidity gap veto, the contribution from DD is expected to be very small and at lower dilepton masses a relatively even mix of SD and elastic production are expected while at higher masses the SD component becomes dominant. Of course additional cuts on e.g. the dilepton acoplanarity can isolate the purely elastic channel further.

We also compare these results with the analytic predictions of~\cite{Harland-Lang:2016apc}, which work in the collinear factorization framework, and model the impact of the rapidity veto by suitably modifying the DGLAP evolution of the photon PDF, i.e. by considering the kinematics of the final--state quark in the $q\to q \gamma^*$ emission and assuming strong DGLAP $k_\perp$ ordering. The survival factor is modelled in a rather similar way to the MC implementation. To compare more directly, we have in fact modified the photon PDF to more closely match the MMHT15 photon~\cite{Harland-Lang:2019pla}, while we also show the effect of varying the factorization/renormalization scale by a factor of 2 around the central value of $\mu = M_{ll}$ to give an estimate of the uncertainty in the prediction in the inclusive case.  We can see that broadly the analytic results are in good agreement with the more precise MC implementation, with the approximate treatment of LO collinear factorization framework and the rapidity veto giving a fair description of the overall trends. Nonetheless we can see that the agreement is not exact, and that there is a reasonable scale variation uncertainty in the theoretical prediction, which is absent in the MC implementation, being based on the structure function approach~\cite{Harland-Lang:2019eai}. Moreover, we emphasise that more differential observables, such as the dilepton transverse momentum/acoplanarity can only be provided by the full MC implementation.

\begin{figure}
\begin{center}
\includegraphics[scale=0.46]{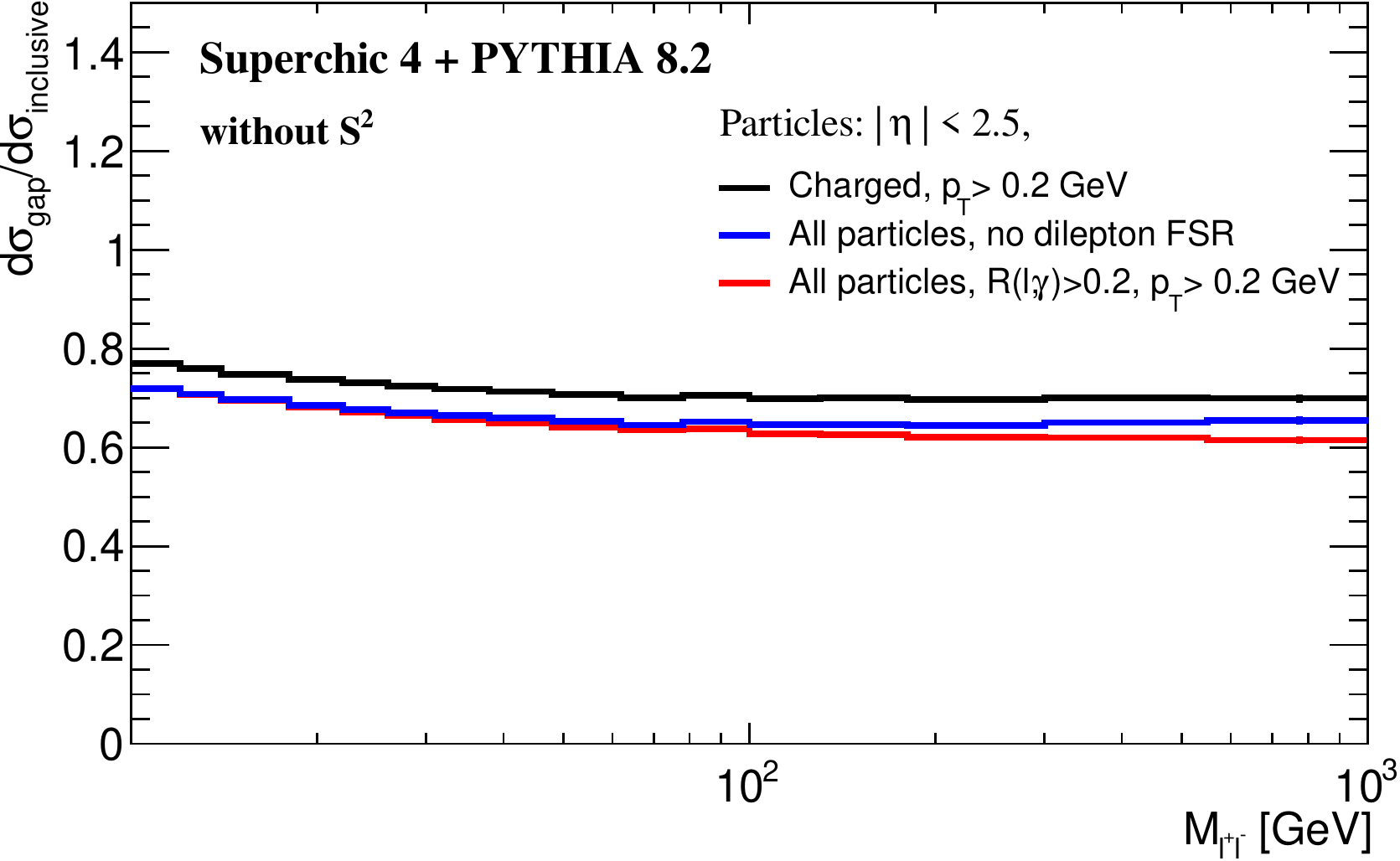}
\includegraphics[scale=0.46]{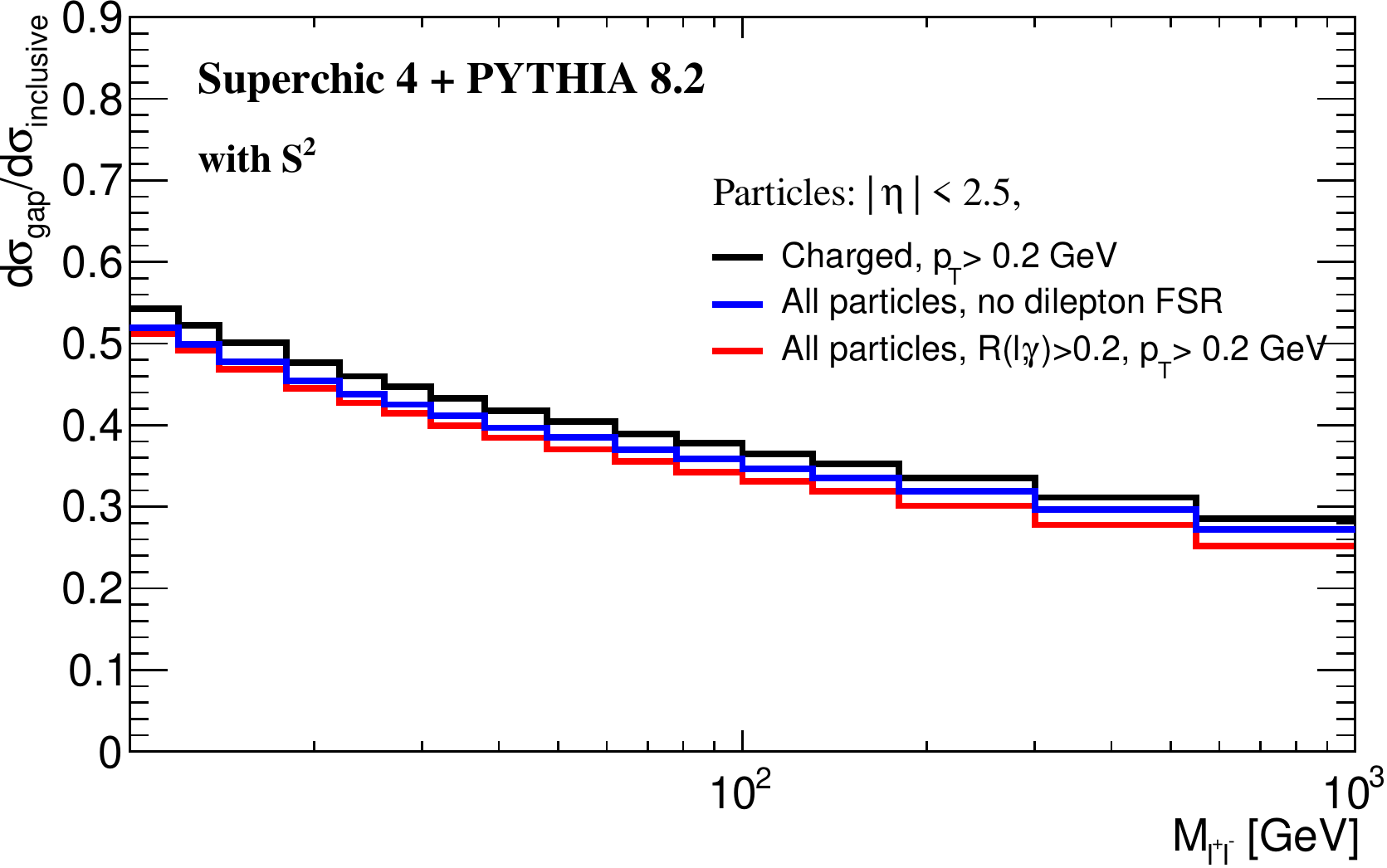}
\caption{The impact of the rapidity veto in the $|\eta|<2.5$ region on the total PI cross section. We show the effect of vetoing on all particles above $p_\perp>0.2$ GeV, allowing photons within a cone of radius $R =0.2$ to be present, as well as vetoing on charged particles only above $p_\perp > 0.2$ GeV. For demonstration purposes we also show the idealised case of a veto on all particles, with no $p_\perp$ threshold and without any photon FSR from the dilepton system. The left (right) figure shows the result excluding (including) the soft survival factor, with it being assumed that all events with additional proton--proton interactions will fail the rapidity veto.}
\label{fig:veto1}
\end{center}
\end{figure} 

Finally, in Fig.~\ref{fig:veto1} we show the impact of the same rapidity veto on the total PI cross section, with the survival factor excluded (included) in the left (right) plot. We show the result of a veto on all particles, which would be relevant for low pile--up running, and with a veto on charged particles which would be relevant for high luminosity running where only the tracker may be used. In both cases we impose realistic $p_\perp$ thresholds and we now take the veto region $|\eta|<2.5$ for concreteness and to allow a direct comparison, though for the all particle case a wider veto region could currently be applied. In the all particle case the situation is somewhat complicated by the presence of FSR photons from the dilepton system, which can lead to the veto on all additional particles being failed. In the case of Fig.~\ref{fig:veto}, this will impact all channels (EL, SD, DD) equally and so will not change their relative contributions, but it will change the overall cross section. To be experimentally realistic, we veto on all photons outside a cone of radius $R=0.2$ around the leptons, though to give an idea of the impact of photon FSR from the dilepton system, we also give for demonstration the result with this switched off in \texttt{Pythia}. In the latter case, we also do not impose any $p_\perp$ threshold, and hence this can be viewed as an idealised situation where we may veto on any additional particles produced at all in the veto region, but not associated with dilepton FSR. 

Considering the broad impact first, we can see that excluding the survival factor the rapidity veto reduces the cross section by $\sim 20\%$, with a relatively mild dependence on the dilepton invariant mass. Including the survival factor we can see that this reduces the cross section further, to $\sim 0.3-0.5$ of the original result, depending on the mass region. For smaller/larger veto regions the reduction will of course be less/more, and this can be readily calculated using the MC for arbitrary scenarios. Now, considering the comparison between the different veto types, we can see that difference between the all particle veto and the charged particle only veto is relatively mild, being $\lesssim 10\%$, and thus a veto on tracks alone is expected to provide quite an accurate evaluation of the true veto, though of course in a realistic analysis one would account for the efficiency of this. Once one imposes a  $p_\perp > 0.2$ GeV threshold and allows photons to lie within a $R=0.2$ radius of the leptons, we can see that the result of this and of simply vetoing on all particles with no threshold and with no FSR photon emission are very similar. If we simply veto on all particles above $p_\perp > 0.2$ GeV then at higher $M_{ll}$ the reduction is  larger.

\subsection{Dilepton acoplanarity distribution: comparison to data}\label{sec:ATLAS}

\begin{figure}
\begin{center}
\includegraphics[scale=0.45]{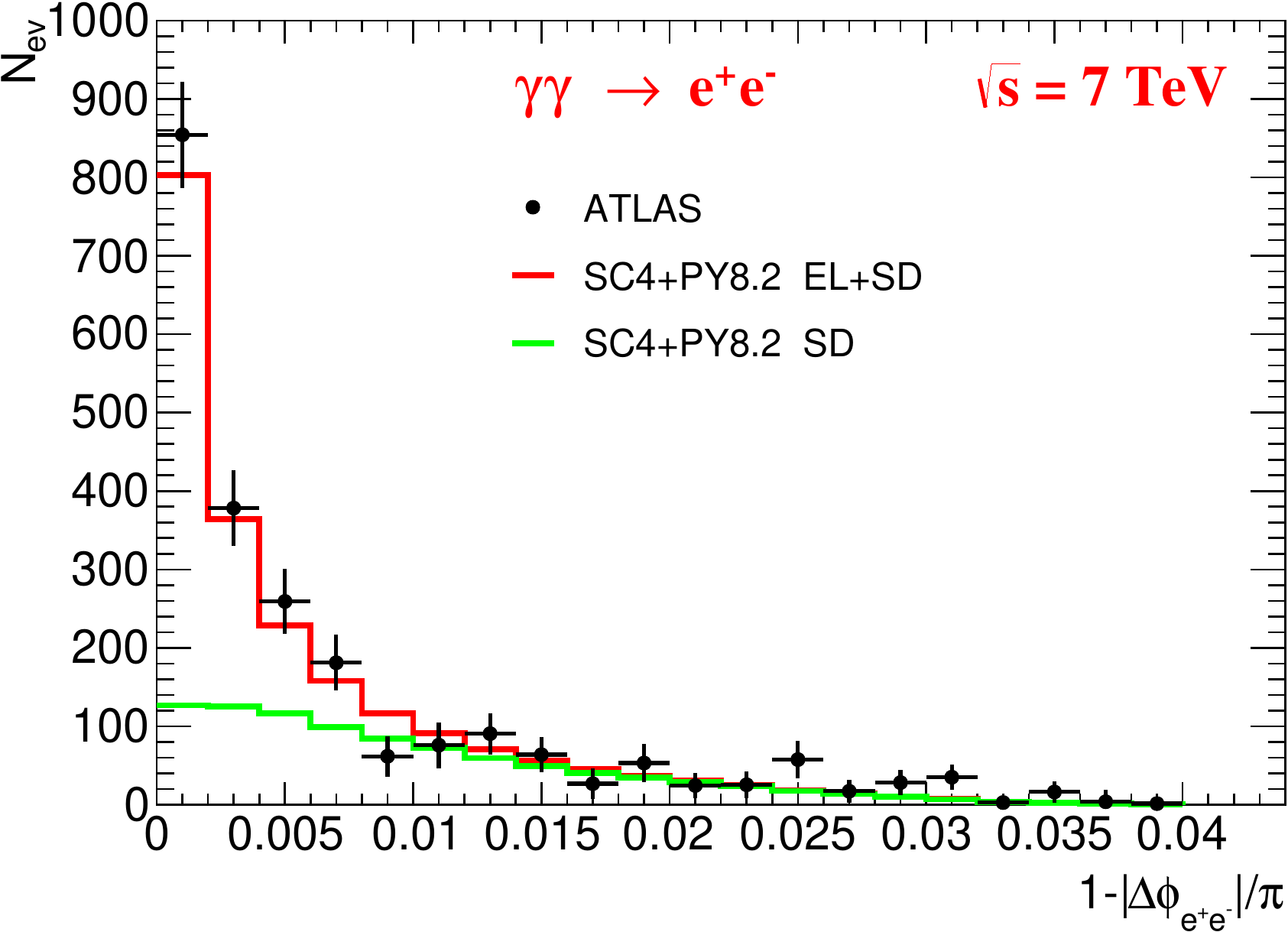}
\includegraphics[scale=0.45]{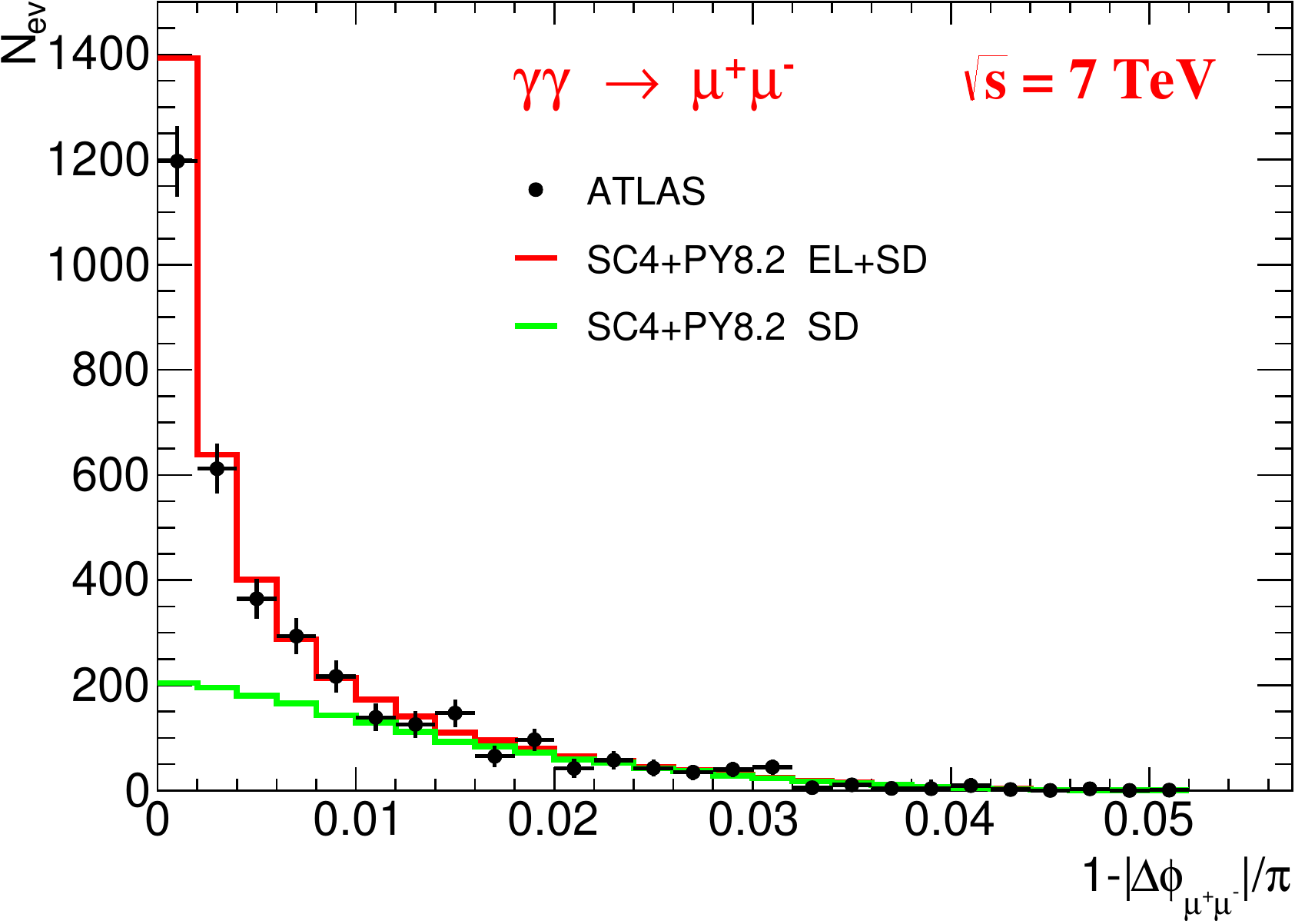}
\caption{Comparison of \texttt{SuperChic 4 + Pythia 8.2} predictions for the dilepton acoplanarity distribution compared to the ATLAS data~\cite{Aad:2015bwa} at $\sqrt{s}=7$ TeV, within the corresponding experimental fiducial region, and with a rapidity veto applied on tracks in the central region. Electron (muon) pair production is shown in the left (right) figures. The elastic and SD contributions are overlaid, while the DD has been subtracted from the data, and so is not included.}
\label{fig:aco}
\end{center}
\end{figure} 

In Fig.~\ref{fig:aco} we compare the predicted acoplanarity distribution for electron (left) and muon (right) pairs to the ATLAS data on semi--exclusive dilepton production at $\sqrt{s}=7$ TeV. This is selected by imposing a veto on additional tracks in association with the dilepton vertex, see~\cite{Aad:2015bwa}  for further details. The Drell--Yan and DD contributions are subtracted from the data, and so we do not include these; we will comment on the latter case further below. We impose the corresponding rapidity veto (although its impact is very
small) directly on our sample of  SD events that were generated without
pile-up, and apply the veto efficiency obtained in the ATLAS analysis evaluated
on samples of elastic events including pile-up to both the elastic and SD events. Pile--up is by far the dominant
effect in reducing the veto efficiency, with values around $\sim 74\%$ for both the
electron and muon channels.
We apply all other cuts on the dilepton system as described in the ATLAS analysis, and in particular a cut on the dilepton $p_\perp^{ll} < 1.5$ GeV, which suppresses the SD contribution and leads to the relatively small impact of the rapidity veto in the absence of pile--up effects. We include the effect of FSR photon emission from the dilepton system.

\begin{figure}
\begin{center}
\includegraphics[scale=0.45]{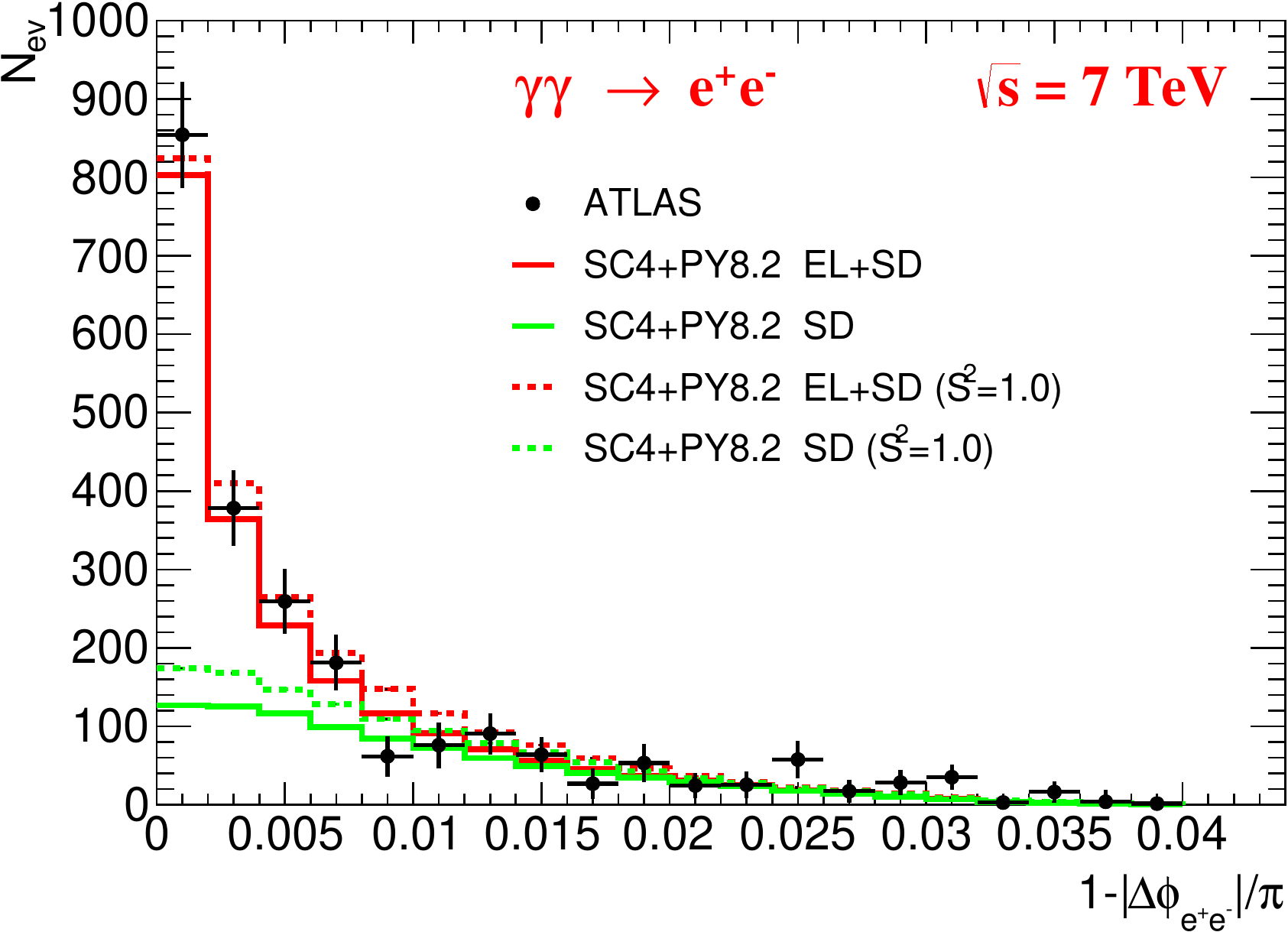}
\includegraphics[scale=0.45]{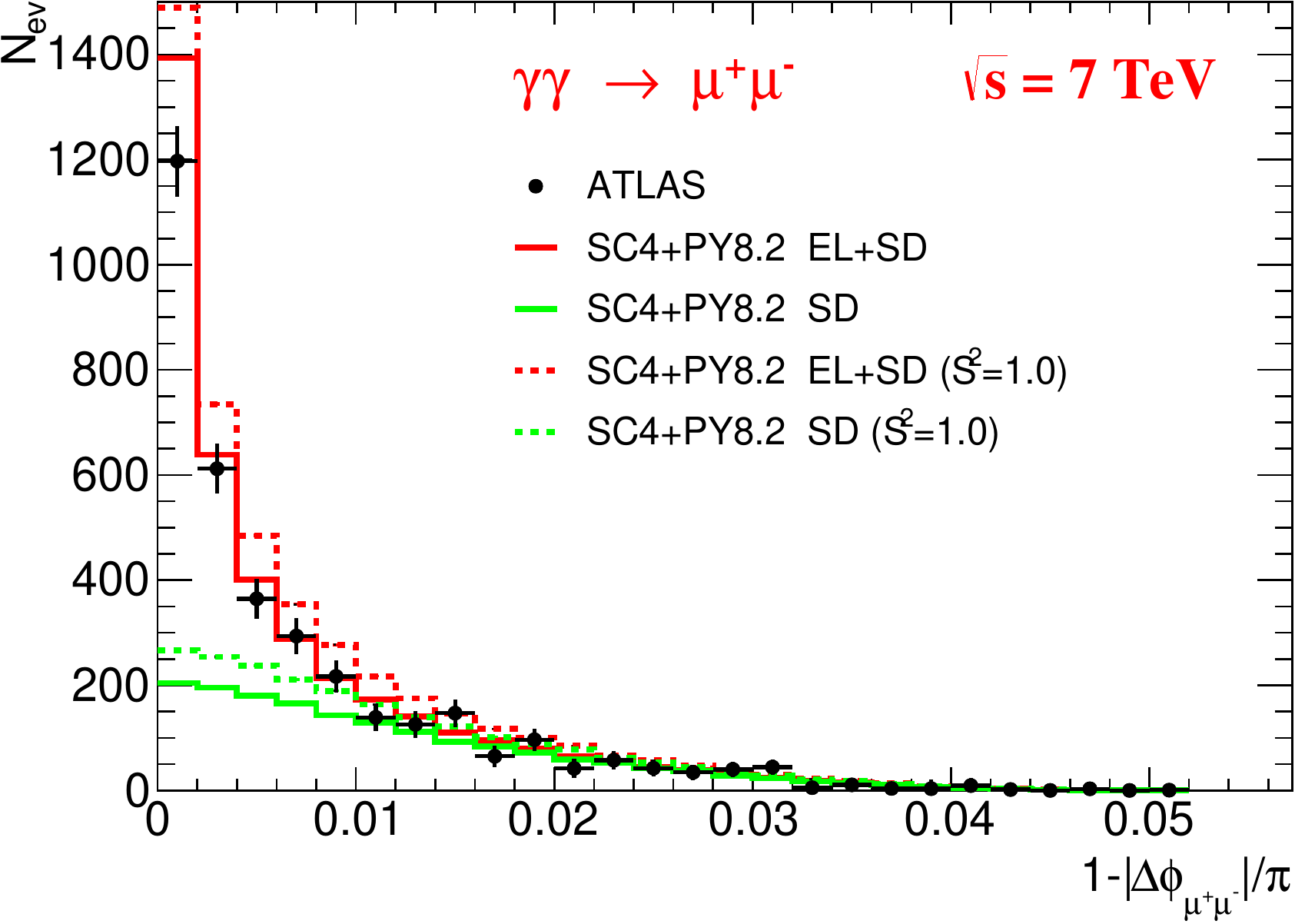}
\caption{As in Fig.~\ref{fig:aco} but with the results without the soft survival factor included shown in addition.}
\label{fig:acoS2}
\end{center}
\end{figure}

The results in the figure are shown overlaid, such that the upper red curve corresponds to the total (elastic + SD) prediction. We can see that the description of the electron data is excellent, and the description of the muon data is generally good. In Fig.~\ref{fig:acoS2} we show the same results, but with the predictions excluding survival effects given in addition, and we can see the importance in including these to achieve a good description of the distributions. On the other hand, in the muon case the predictions appear to overshoot the measurement in the lowest acoplanarity bin somewhat, where the elastic contribution is enhanced. Given the relatively limited statistics and apparent mild inconsistency between the two samples, for which the $p_\perp^l$ cuts are slightly different but otherwise the fiducial region is the same, it is difficult to make a firm statement about this. However, the statistics are higher in the muon sample and so this is certainly a question that requires clarification. Indeed, in the ATLAS 13 TeV measurement~\cite{Aaboud:2017oiq}, although not presented in a form that we can compare our differential results directly to, the {\tt SuperChic} predictions (albeit with an earlier version of the MC) are also found to overshoot the extracted elastic component of the data somewhat. Some tuning of the modelling of the survival factor could for example improve this agreement, or the apparent discrepancy may lie with the data itself.

Future data, in particular presented differentially and in regions where the elastic contribution is both enhanced and suppressed, should help clarify the situation. Nonetheless, we find these first results encouraging, and emphasise that in the SD case this corresponds to a prediction of the MC, whereas in the ATLAS analysis the SD contribution is modelled via \texttt{LPAIR 4.0}, which does not account for survival effects at all, with the normalization fixed to the data. We consider our MC implementation to be a significant advance with respect to such an approach.

A further useful element will be to compare against data for which the DD has not been subtracted. In the ATLAS analysis~\cite{Aad:2015bwa} this is evaluated using LO photon--initiated production calculated within collinear factorization, as implemented in \texttt{PYTHIA 6.425}~\cite{Sjostrand:2006za} and with the NNPDF2.3QED~\cite{Ball:2013hta} PDF set. As the PDF set used in~\cite{Aad:2015bwa} is in fact rather outdated, and this relies on a purely LO calculation, the corresponding prediction is certainly not guaranteed to be accurate. This is not expected to be a significant issue from the point of the view of the comparisons in Fig.~\ref{fig:aco}, as the expected DD contribution is rather small. Nonetheless, it is worth comparing our prediction for the DD contribution to this. After applying all cuts, we find that the expected ratio of DD to SD events is $\approx 10\%$ for both channels, whereas in the ATLAS analysis a larger value of $\approx 15-20\%$ is found; the acoplanarity distributions are similar on the other hand. Such a difference may in part be explained by the differing treatment of survival effects, which in the ATLAS analysis are included via the default \texttt{Pythia} MPI model. Indeed, we observe from the ATLAS results that the impact of the rapidity veto is predicted to give roughly a $\approx 20\%$ reduction in the DD case, which will largely be driven by the inclusion of MPI. This is clearly a smaller reduction than the predicted survival factor in the DD case, see Fig.~\ref{fig:S2}. However, we note that this is before placing a cut on the dilepton $p_\perp^{ll} < 1.5$ GeV, which we find impacts the role of survival effects within our approach considerably (as it cuts out the larger $Q^2$ region where the survival factor is lowest) and moreover the LO cross section calculation with the outdated NNPDF2.3QED may well differ from our prediction rather significantly.

In summary, though our initial results are encouraging, it would be very interesting to compare against data selected in regions where the elastic SD, and DD enter, without subtracting any dissociative contributions, such that our approach can be tested more precisely in all channels.

We finish this section by noting in the ATLAS 8~\cite{Aaboud:2016dkv} and 13 TeV~\cite{Aaboud:2017oiq} analyses the data are compared against the `finite size' predictions of~\cite{Dyndal:2014yea}, for which the survival factor is somewhat lower than the predictions from \texttt{SuperChic}. It is important to emphasise again, as we have discussed already in~\cite{Harland-Lang:2015cta}, that this approach for calculating the survival factor is based on the same physical principles as here, but with a simplified form for the proton opacity. The reason for this lower survival factor prediction is highly unlikely to be due to this difference, however, but is rather due to the fact that in~\cite{Dyndal:2014yea} the spin structure of the $\gamma \gamma \to l^+ l^-$ amplitude is completely (and incorrectly) omitted, and an additional  cut is placed on the impact parameter $b_t$ between the lepton pair production point and the centre of each colliding proton, $b_t > r_p$, where $r_p$ is the proton radius, irrespective of the separation between the protons themselves. This latter cut is certainly unjustified, and effectively assumes that the lepton pair and the protons may interact strongly to fill the rapidity gap. It will by definition reduce the predicted survival factor in an unphysical way. This point is also discussed in~\cite{Azevedo:2019fyz} in the context of heavy ion collisions. Until these issues are corrected, one cannot meaningfully compare the results of this model with the data. Indeed, when and if they are corrected, we expect the corresponding predictions to lie closer to those presented here. We note that a similar such unphysical cut is imposed by default in the \texttt{STARLIGHT} MC~\cite{Klein:2016yzr} for the case of lepton pair production in heavy ion collisions.

\subsection{Dark matter Searches in Compressed Mass Scenarios: Background Evaluation}\label{sec:DM}

A further useful application of our work relates to the study outlined in~\cite{Harland-Lang:2018hmi}, which examines the possibility of observing SUSY particles with EW couplings in a compressed mass scenario, with the signal corresponding to relatively low $p_\perp$ leptons from the decay of the SUSY particles and FPD tags for the elastic protons. To be precise, the pair production of smuons and selectrons $\tilde{l}_{L,R}$ ($l=e,\mu$), where $L$ and $R$ denotes the left and right handed partner of the electron or muon, was considered. The four sleptons were taken to be mass degenerate and to decay with 100\% branching ratio to the corresponding SM partner, and an invisible neutralino, $\tilde{\chi}_1^0$. Slepton masses in the range 120-300 GeV were considered.

While  elastic lepton pair production represents a negligible background, as the required dilepton invariant mass reconstructed from the FPD tags would correspond to much higher dilepton invariant masses, a potential background is from SD or DD lepton pair production at relatively low lepton transverse momenta, where a proton from the dissociation system(s) is tagged in the FPD and the missing mass reconstructed from the corresponding proton fractional momentum loss, $\xi$, values passes the event selection. The probability for this to occur from a given dissociation vertex is certainly low, but given the large cross section for low mass dilepton production, this can still be an important background. 

In~\cite{Harland-Lang:2018hmi}, various cuts were imposed on the dilepton system as well as the tagged protons in order to reduce this background. In particular, a transverse momentum cut of $p_{\perp,{\rm proton}} < 0.35$ GeV was placed, as protons from the dissociation system tend to be produced at higher $p_{\perp}$ than elastic protons, while cuts to select larger dilepton acoplanarity and transverse momentum difference were applied in order to enhance the signal in comparison to the background. In the latter case, while the leptons from SUSY particle decays will be produced at arbitrary  acoplanarity, leptons in PI production will be relatively back--to--back, though this will be washed out somewhat by the recoil from the proton dissociation system(s) in the SD(DD) cases. To get a firm handle on the impact of such cuts on the background clearly requires a precise differential knowledge of the kinematics of the dilepton system. This was not yet available in~\cite{Harland-Lang:2018hmi}, and hence here instead we had to rely on a rather approximate treatment, based on the approach of~\cite{Harland-Lang:2016apc}, whereby the DGLAP evolution of the collinear photon PDF was modified to effectively account for these cuts. 

The result from this approach was that the corresponding background could be reduced to $\sim 1$ expected event or less, and would therefore be rather low. However, the expected signal itself only corresponded to a handful of events, and therefore a more precise evaluation of this background is clearly in order. We now have the tools to do this, with our MC implementation allowing a fully differential treatment of the dilepton system and hence determination of the impact of the above cuts on this background. Moreover, the particle--level treatment of the proton system automatically allows us to include the probability for protons from the dissociation to fall within the FPD acceptance, including the effect of the $p_{\perp,{\rm proton}}$ cut.

Our results are show in Table~\ref{mumu}. We can see that, as in~\cite{Harland-Lang:2018hmi} the SD background is completely negligible, with $\sim 0.01$ or less events expected in all cases. The DD is larger, but still very small, with less than 0.5 events expected in all cases. We note that in the approximate estimate of~\cite{Harland-Lang:2018hmi} the expected number of DD events was higher, being $\sim 1.4$ (1.1) in the  $|\eta|<2.5$ ($|\eta|<4.0$) regions. Our more precise results are  lower by a factor of $\sim 3$ ($10$) in the  $|\eta|<2.5$ ($|\eta|<4.0$) regions. For the larger $\eta$ coverage we expect the previous estimates, which do not explicitly account for the dilepton rapidity dependence, to be most approximate, and hence it is interesting to see that the difference is largest in this case.
We therefore expect this background to be very low, and certainly less significant than our previous results suggested. Of course, in a full experimental analysis one would not necessarily rely on these predictions alone, but rather could measure the contributions from lepton pair production in a broader phase space region in order to control for it in the signal region. For such a study, a precise MC simulation will be essential, and is provided here.

\begin{table}\begin{center}
\begin{tabular}{|c|c|c|} 
\hline
&  $|\eta|<2.5$  &  $|\eta|<4.0$ \\
\hline
SD & 0.01 & $\sim 0$ \\ \hline
DD &0.4 & 0.1 \\ \hline
\end{tabular}
\caption{\small{Event yields for an integrated luminosity of 300 ${\rm fb}^{-1}$ for 
  dilepton production  after applying all cuts
  specified in Table 2 of~\cite{Harland-Lang:2018hmi}. Results for single and double proton dissociation are given, and with
different pseudorapidity intervals for both the final--state leptons and rapidity veto. Lepton
  reconstruction efficiencies are taken from~\cite{Aaboud:2017leg}. The values marked as $\sim$0
correspond to numbers which are sufficiently below 0.01.}}
    \label{mumu}
    \end{center}
\end{table}

 \section{Summary and Outlook}\label{sec:conc}

In this paper we have presented the results of a new MC implementation of PI production in proton--proton collisions. This is released in the \texttt{SuperChic 4} MC, which as well as the case of PI dilepton production discussed here can generate a range of other processes, as described in~\cite{Harland-Lang:2015cta,Harland-Lang:2018iur}. The code and a user manual can be found at 
\\
\\
{\tt http://projects.hepforge.org/superchic}
\\
\\
The results presented in this work are based on the structure function calculation of the underlying production process, which as well as providing a precise prediction for the inclusive cross section, allows the different channels where either (single dissociation), both (double dissociation) or neither (elastic) proton interacts inelastically and dissociates, to be evaluated individually. The MC events for these channels are then interfaced to {\tt Pythia} for showering and hadronisation.
This is essential for the modelling of PI production with rapidity gaps in the final state, which are naturally produced due to the colour--singlet nature of the photon. Such events are commonly selected by imposing a veto on additional particle production, and in this case we must account for both the possibility that the proton dissociation products fill the veto region, as well as that additional proton--proton interactions may occur, filling the gap with MPI activity. The probability that the latter does not occur is often referred to as the soft survival factor. Both of these effects, and their interplay with the specific kinematics of the process, have been fully included in our calculation. The case that elastic protons are tagged in FPDs is in addition modelled within such an approach.

We have presented predictions for the different contributions of the elastic and proton dissociative channels to the total cross section, with and without a rapidity veto imposed. Such a veto is found to have a large impact on the results, naturally enhancing the elastic component. We have also presented results for the soft survival factor. Broadly speaking, for elastic production the average proton impact parameter is rather large and hence the impact of additional proton--proton interactions rather low, that is the survival factor is quite close to unity. This remains true for single dissociation, while for double dissociation the survival factor is expected to be significantly lower. Other more subtle dependencies of the survival factor on the final--state kinematics have also been mapped out in detail.

We have in addition compared our results to the ATLAS measurement of semi--exclusive electron and muon pair production at 7 TeV, finding our predictions for both the cross section normalizations and acoplanarity distribution to be generally in good agreement with the data. Finally, we have presented updated predictions for the background from semi--exclusive dilepton production to slepton pair production in a compressed mass scenario. Our MC implementation allows this to be evaluated for the first time with precision, and we find that the expected background is significantly smaller, by a factor of $\sim 3-10$ in comparison to earlier estimates.

While we have considered in detail the case where a rapidity veto is imposed on the final state, we note that the MC also provides a high precision prediction for inclusive PI production, simply by summing over the elastic, SD and DD channels and omitting the survival factor. A detailed analysis of this will be presented in future work.
In addition, while our study focussed on $pp$ collisions, we note that this approach could be straightforwardly applied for PI production in $pA$ collisions, where any pile--up background is absent.
In the case of $pp$ collisions, there is already an active ongoing and future programme of measurements for exclusive and semi--exclusive PI production, both with and without tagged protons, during nominal LHC running. An essential element of this is a comprehensive MC implementation of the underlying process, which we provide for the first time here. Though we have only discussed the case of lepton pair production in the current work, we emphasise that this MC can be readily extended to the PI production of other SM and BSM states. This will be the focus of future studies.

\section*{Acknowledgments.}

We are grateful to Ilkka Helenius for a significant amount of useful discussion and feedback about this work. We are grateful to Ilkka Helenius and  Torbj{\"o}rn Sj{\"o}strand  for help in interfacing to \texttt{Pythia}. We are grateful to Mateusz Dyndal, Peter Steinberg and Radek Zlebcik for useful discussions.
LHL thanks the Science and Technology Facilities Council (STFC) for support via grant award ST/L000377/1. MT is supported by MEYS of the Czech Republic within
project LTT17018.

\bibliography{references}{}
\bibliographystyle{h-physrev}

\end{document}